# Enhanced Confocal Laser Scanning Microscopy with Adaptive Physics Informed Deep Autoencoders

Zaheer Ahmad[1], Junaid Shabeer[2], Usman Saleem[3], Tahir Qadeer[4], Abdul Sami[5], Zahira El Khalidi[6], Saad Mehmood[7]

[1]Department of Physics & Astronomy, Georgia State University, Atlanta, GA, 30303, USA
[2] Department of Physics, Riphah International University, Islamabad, 46000, Pakistan
[3]Roots IVY, Islamabad, 44000, Pakistan
[4,5]Arid Agriculture University, Rawalpindi, 46000, Pakistan
[6]Department of Physics, University of Illinois, Chicago, IL, 60607, USA
[7]Department of Physics, University of Central Florida, Orlando, FL, 32816, USA

**Abstract**

We present a physics-informed deep learning framework to address common limitations in Confocal Laser Scanning Microscopy (CLSM), such as diffraction limited resolution, noise, and undersampling due to low laser power conditions. The optical system's point spread function (PSF) and common CLSM image degradation mechanisms namely photon shot noise, dark current noise, motion blur, speckle noise, and undersampling were modeled and were directly included into model architecture. The model reconstructs high fidelity images from heavily noisy inputs by using convolutional and transposed convolutional layers. Following the advances in compressed sensing, our approach significantly reduces data acquisition requirements without compromising image resolution. The proposed method was extensively evaluated on simulated CLSM images of diverse structures, including lipid droplets, neuronal networks, and fibrillar systems. Comparisons with traditional deconvolution algorithms such as Richardson-Lucy (RL), non-negative least squares (NNLS), and other methods like Total Variation (TV) regularization, Wiener filtering, and Wavelet denoising demonstrate the superiority of the network in restoring fine structural details with high fidelity. Assessment metrics like Structural Similarity Index (SSIM) and Peak Signal to Noise Ratio (PSNR), underlines that the *AdaptivePhysicsAutoencoder* achieved robust image enhancement across diverse CLSM conditions, helping faster acquisition, reduced photodamage, and reliable performance in low light and sparse sampling scenarios holding promise for applications in live cell imaging, dynamic biological studies, and high throughput material characterization.

## 1. Introduction

Recent progresses in microscopy needs higher imaging speeds to accommodate dynamic biological studies, three dimensional reconstructions, and high throughput experiments.[i,ii,iii,iv,v,vi,vii] On the other hand, resolution enhancements often result in increased noise due to fewer photons being available per pixel, especially in low illumination conditions. These challenges have accelerated the search for methods to enable efficient data acquisition without compromising image quality.

Confocal Laser Scanning Microscopy (CLSM) offers high resolution imaging with optical sectioning and detailed three-dimensional reconstructions.[viii,ix,x,xi,xii,xiii,xiv] However, under practical conditions, CLSM images are often degraded by factors such as diffraction-limited resolution, photon shot noise, motion blur, and undersampling artifacts. The optical Point Spread Function (PSF)[xv,xvi,xvii,xviii,xix], inherent to the system, limits spatial resolution and causes blurring. Furthermore, dark current and speckle noise degrade CLSM image quality further to the point that maintaining clarity in low light and/or power sensitive conditions becomes challenging. Undersampling or non-uniform illumination, further worsen these issues.

Traditional approaches to improving CLSM image quality rely on post-acquisition techniques such as deconvolution and denoising, which often address the symptoms of degradation but are unable to incorporate the underlying physical principles of image formation. Hardware-based solutions, including advanced detectors or adaptive optics, provide direct improvements but are often expensive and not universally accessible.

Deep learning has proven effective for image restoration, demonstrating the ability to learn mappings between degraded and high-quality images.[xx,xxi,xxii,xxiii] However these methods are usually designed as general tools without leveraging specific physics of the imaging process. Physics-informed models address this gap by modeling physics of optical systems and noise mechanisms into the model training process, allowing the network to achieve better consistency with the physical constraints of imaging.

In the current work, we developed a physics-informed autoencoder for CLSM image restoration that incorporates the imaging system's point spread function (PSF), diffraction effects, noise mechanisms, and sampling constraints directly into its design. This study is to focused on addressing the imaging challenges due to noisy conditions in CLSM of various biosystems, where prolonged light exposure and/or high intensity lasers can potentially damage or, in some cases, destroy the samples. This imposes imaging under low light and low laser power conditions to preserve sample integrity. However, these conditions result into undersampling, increased noise levels, and reduced image resolution. Therefore, integrating such physical constraints into the autoencoder we aim to find a robust solution for restoring high-quality CLSM images while ensuring minimal impact on delicate biological samples. The model simulated common CLSM imaging degradations, including photon shot noise, motion blur, speckle noise, and undersampling, and used these to train a physics constrained neural network that restores high quality CLSM images from degraded inputs. Our methods reduces the need for additional hardware or modifications to existing CLSM systems, focusing instead on software based solutions to improve image quality in CLSM. The proposed method is tested on varied synthetic CLSM datasets, demonstrating its capability to restore spatial resolution and reduce noise while maintaining consistency with the physical principles of CLSM imaging.

## 2. Common Noise types in Confocal Laser Scanning Microscopy

Confocal Laser Scanning Microscopy (CLSM) imaging faces inherent limitations due to physical degradations caused by optical diffraction, noise sources, and undersampling. These degradations are amplified under conditions of low laser power, which is essential to prevent photodamage to delicate samples.

The imaging resolution in CLSM is dictated by the diffraction of light, represented by the Point Spread Function (PSF)[xxiv]. The PSF describes the intensity distribution of light in the focal plane, determined by the Airy disk pattern[xxv,xxvi]:

$$PSF(r) = [\frac{2J_1(\pi D \frac{r}{\lambda})}{\pi D \frac{r}{\lambda} + \epsilon}]^2$$

Such that, $J_1$ is first order Bessel function, D is numerical aperture, r is the radial distance in focal plane, $\lambda$ is laser wavelength and $\epsilon$ is an arbitrary constant to avoid division by zero.

The lateral resolution of CLSM is given by[xxvii,xxviii]

$$d = \frac{0.61\lambda}{D}$$

Similarly, the axial resolution (which defines depth discrimination) is given as

$$d_z = \frac{2\lambda}{nD^2}$$

There are several other noise sources that degrade CLSM images. Photon shot noise arises from the quantum nature of light, where the number of detected photons fluctuates statistically. Light intensity affected by photon shot noise is modeled as

$$I_{shot}(x,y) = I(x,y) + R_{random}\sqrt{I(x,y)}$$

such that, $I(x,y)$ is true intensity for a pixel at $(x,y)$ and $R_{random}$ is a Gaussian random variable with zero mean and unit variance.

Similarly, there is a also dark current noise, that originates from thermally excited electrons in the detector. The intensity of light $I(x,y)$ for a pixel at $(x,y)$ affected by dark current noise is expressed mathematically as

$$I_{dark}(x,y) = I(x,y) + R_{dark}\sigma_{dark}$$

Speckle noise, caused by coherent light interference, is modeled for the same pixel as multiplicative noise, with $\sigma_{speckle}$ as speckle noise strength:

$$I_{speckle}(x,y) = I(x,y)(1 + R_{speckle}\sigma_{speckle})$$

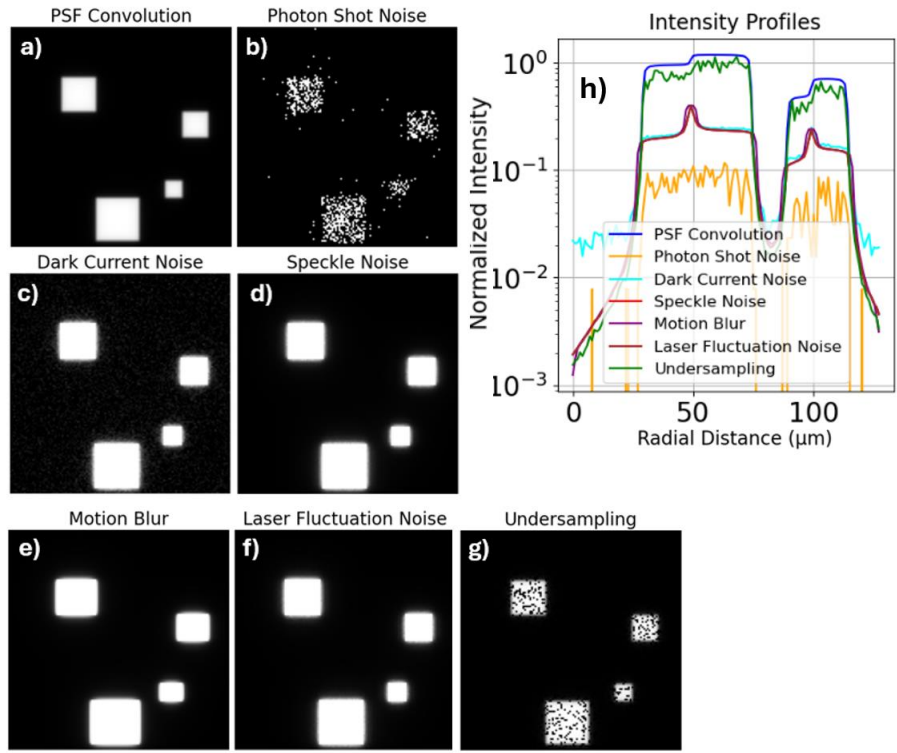

*Figure 1: Common degradation mechanisms in Confocal Laser Scanning Microscopy (CLSM). (a) Image degraded by the Point Spread Function (PSF), representing optical diffraction and blurring. (b) Photon shot noise, caused by statistical fluctuations in photon detection. (c) Dark current noise, originating from thermally excited electrons in the detector. (d) Speckle noise, due to coherent light interference. (e) Motion blur, introduced by sample drift or mechanical vibrations. (f) Laser fluctuation noise, caused by instability in laser power. (g) Undersampling artifacts, resulting from insufficient pixel density or random pixel omission. (h) Intensity profiles corresponding to each noise type, showing their impact on the normalized intensity along the radial distance.*

Sample drift or any kind of mechanical vibrations introduce motion blur in CLSM image, and it can be mathematically modeled using a blur Kernal as a convolution "*" below:

$$I_{blur}(x, y) = K_{blur} * I(x, y)$$

Any fluctuations in laser power may also introduce noise in CLSM images, and it can be expressed as

$$I_{fluctuation}(x, y) = I(x, y)(1 + R_{fluctuation}\sigma_{fluctuation})$$

And the undersampling of a CLSM image can be simulated using a binary mask $M(x, y)$ with pixel omission probability p is defined as

$$M = \begin{cases} 1 & ; probabliity = 1 - p \\ 0 & ; \phantom{xx} probability = p \end{cases}$$

And the affected intensity will then be

$$I_{undersampled}(x, y) = I(x, y) \cdot M(x, y)$$

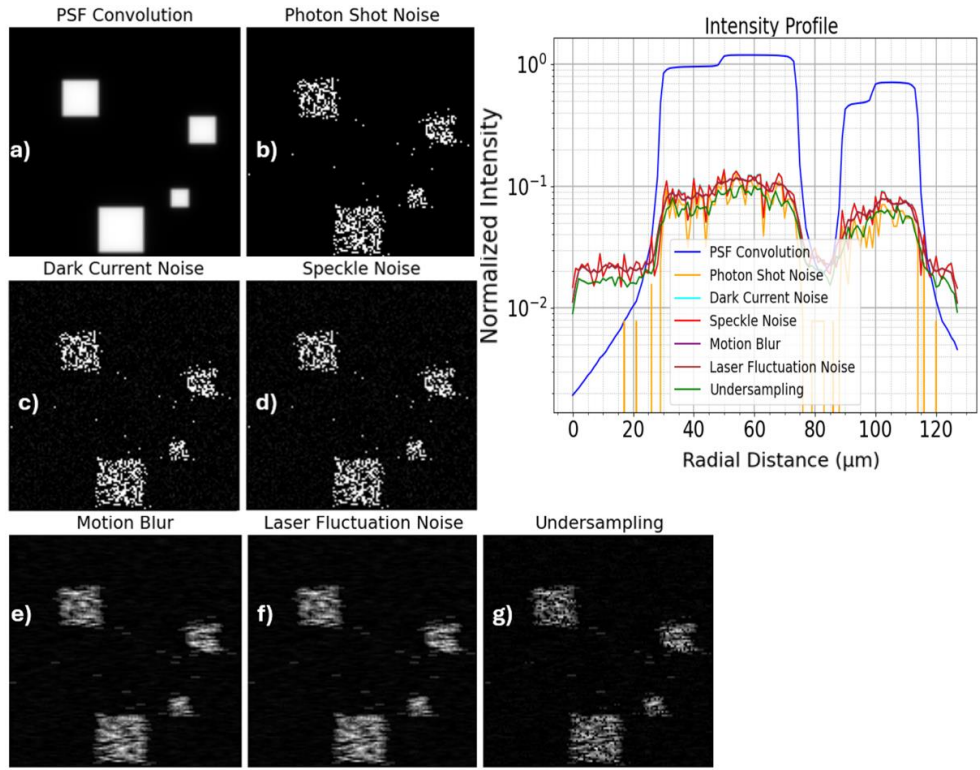

***Figure 2: Cumulative effects of noise types in CLSM imaging,** each subsequent image has added a new noise type to the previous ones: (a) PSF Convolution, (b) + Photon Shot Noise, (c) + Dark Current Noise, (d) + Speckle Noise, (e) + Motion Blur, (f) + Laser Fluctuation Noise, and (g) + Undersampling. Right Panel: Normalized intensity profiles showing the combined impact of progressively added noise types on image quality. **Right Panel**: The normalized intensity profiles corresponding to the noise types, showing their radial intensity distribution across the CLSM image. Each noise type exhibits distinct intensity variations and degradation effects.*

The individual effect of each of the noise types on a given image is displayed in fig. 1 with their effect on intensity profile displayed in fig 1h. However, real life CLSM imaging may contain a combination of few or all such noise types. Fig 2. displays sequential degradation of an example image.

Heat is also generated during CLSM imaging due to interaction of Laser with the sample being scanned. Diffusion equation describes the heat $q$ absorbed by sample with specific heat $c_p$ density $\rho$ and thermal diffusivity $\alpha$, at temperature $T$ for time interval $t$

$$\frac{\partial T}{\partial t} = \alpha \nabla^2 T + \frac{q}{\rho c_p}$$

$$q = \mu_{abs} P \exp(-\mu_{abs} z)$$

; $\mu_{abs}$, $P$ and $z$ are absorption coefficient, laser power, and penetration depth, respectively.

## 3. Methods: Adaptive Physics Autoencoder

To mitigate these degradations, we developed an Autoencoder model, whose training was constrained by the physics confocal laser scanning microscopy and all those noise types described in the previous section. An autoencoder is a type of neural network with an encoder decoder structure. The encoder maps noise degraded image X into a compressed latent space Z

$$Z = f_{encoder}(X; \theta)$$

The latent space representation $Z$ encodes the image while filtering out noise. It follows the constraints imposed by the Point Spread Function (PSF) defined previously by equation (x), which governs the spatial resolution.

The decoder then reconstructs a clean denoised image from Z.

$$\hat{X} = f_{decoder}(Z; \phi)$$

The $\theta$ and $\phi$ are encoder and decoder parameters, respectively. The training objective minimizes the Mean Squared Error (MSE) between reconstructed images X and ground truth images $\hat{X}$. The primary loss term, the Mean Squared Error (MSE), is defined for N training samples as

$$\mathcal{L}_{MSE} = \frac{1}{N} \sum_{i=}^{N} \left\| \hat{X}^{(i)}(x,y) - X^{(i)}(x,y) \right\|^2$$

To ensure the reconstructed image adheres to the photon conservation law in CLSM, a photon budget loss between reconstructed and ground truth images is introduced

$$\mathcal{L}_{photon} = \left| \sum \hat{X} - X \right|$$

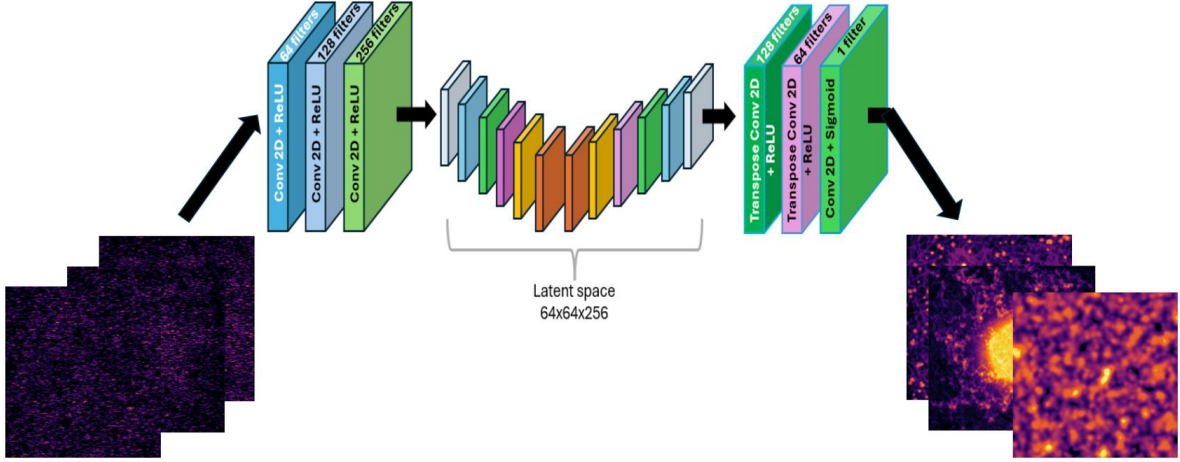

*Figure 3: Deep learning architecture for image restoration in Confocal Laser Scanning Microscopy (CLSM). The model is a convolutional autoencoder comprising an encoder (left) and a decoder (right) with a latent space of size 64×64×256. The encoder, built with 2D convolutional layers and ReLU activation, extracts features from noisy input images. The decoder reconstructs the restored images using transposed convolutions, ReLU activations, and a final sigmoid activation to produce high-resolution outputs.*

The morphological preservation of features in CLSM images was ensured by edge loss defined as

$$\mathcal{L}_{edge} = |\nabla \hat{X} - X|$$

And the total loss function for training was defined to be linear combination of the above defined losses, with hyperparameters $\lambda_1$ and $\lambda_2$ to control relative contributions of various loss types.

$$\mathcal{L}_{total} = \mathcal{L}_{MSE} + \lambda_1 \mathcal{L}_{photon} + \lambda_2 \mathcal{L}_{edge}$$

The training data was augmented with simulated noise types defined in previous section to accommodate for real life noisy conditions encountered in confocal laser scanning microscopy.

The training progression of the Adaptive Physics Autoencoder for restoring CLSM images is shown in Figure 4. Starting with the degraded input i.e. Network Input, the reconstructed images progressively converge toward the Ground Truth with gradual recovery of fine structural details as training progresses. This is also complemented quantitatively by normalized intensity profiles that show improved fidelity in reconstructions, with closer alignment to the Ground Truth over successive epochs. Quantitative metrics i.e. Structural Similarity Index (SSIM) and Peak Signal-to-Noise Ratio (PSNR) also indicate consistent improvements in image quality throughout training, achieving optimal performance at 300 epochs.

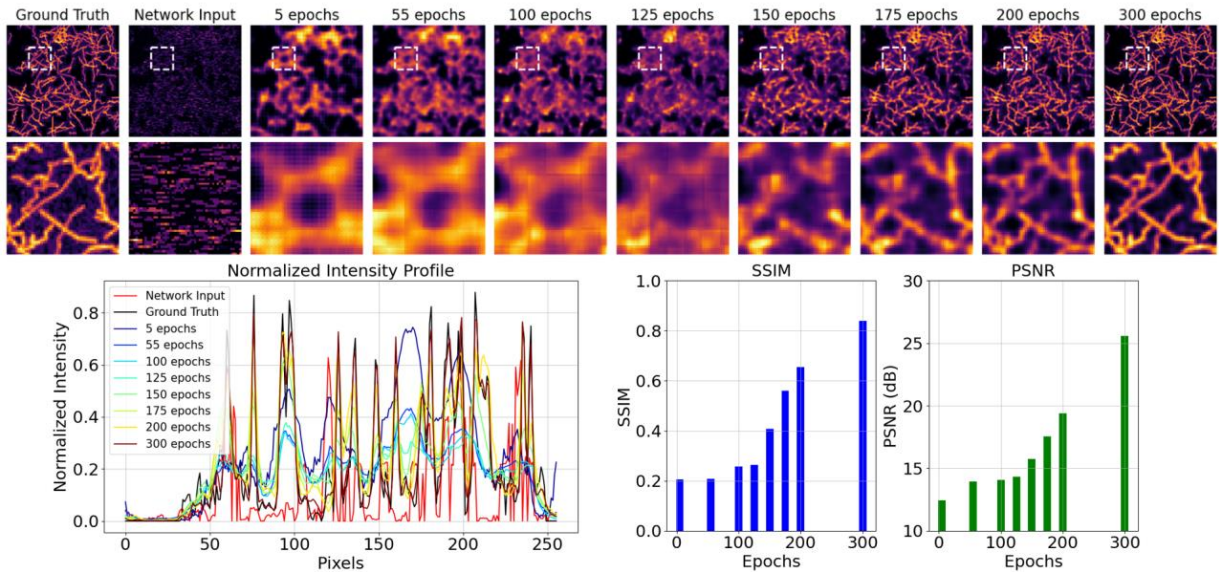

*Figure 4: Training progression of the proposed Adaptive Physics Autoencoder for Confocal Laser Scanning Microscopy (CLSM) image restoration. The top row shows reconstructed images at different epochs, starting from degraded input (Network Input) and progressively approaching the Ground Truth with increasing training epochs. Zoomed-in Regions of Interest (ROIs) highlight the gradual recovery of fine structural details in the reconstructions. The normalized intensity profiles (bottom left) compare the fidelity of the reconstructions at various epochs against the Ground Truth within the ROI, demonstrating increasing alignment as training progresses. Quantitative metrics including Structural Similarity Index (SSIM) and Peak Signal-to-Noise Ratio (PSNR) (bottom middle and right) show steady improvements in image quality, reaching optimal performance at 300 epochs.*

## 4. Results and Discussion
### 4.1 Performance on CLSM images of lipid droplet morphology in adipocytes

We first demonstrate the ability of the proposed Adaptive Physics Autoencoder to improve the degraded CLSM images of lipid droplets in a gel matrix. Figure 5 shows the visualization of the recovery performance of different algorithms. The network output images are inferred from the degraded images, i.e., the network input image. We compare the visualized reconstruction performance of the proposed network with widely used image deconvolution algorithms, including the non-negative least squares (NNLS) algorithm and Richardson-Lucy (RL) algorithm.

For a fair comparison, the degraded images were up-sampled by bicubic interpolation before being deconvolved. It can be seen that the resolution of the images processed by the two deconvolution algorithms is somewhat improved compared to the input images.

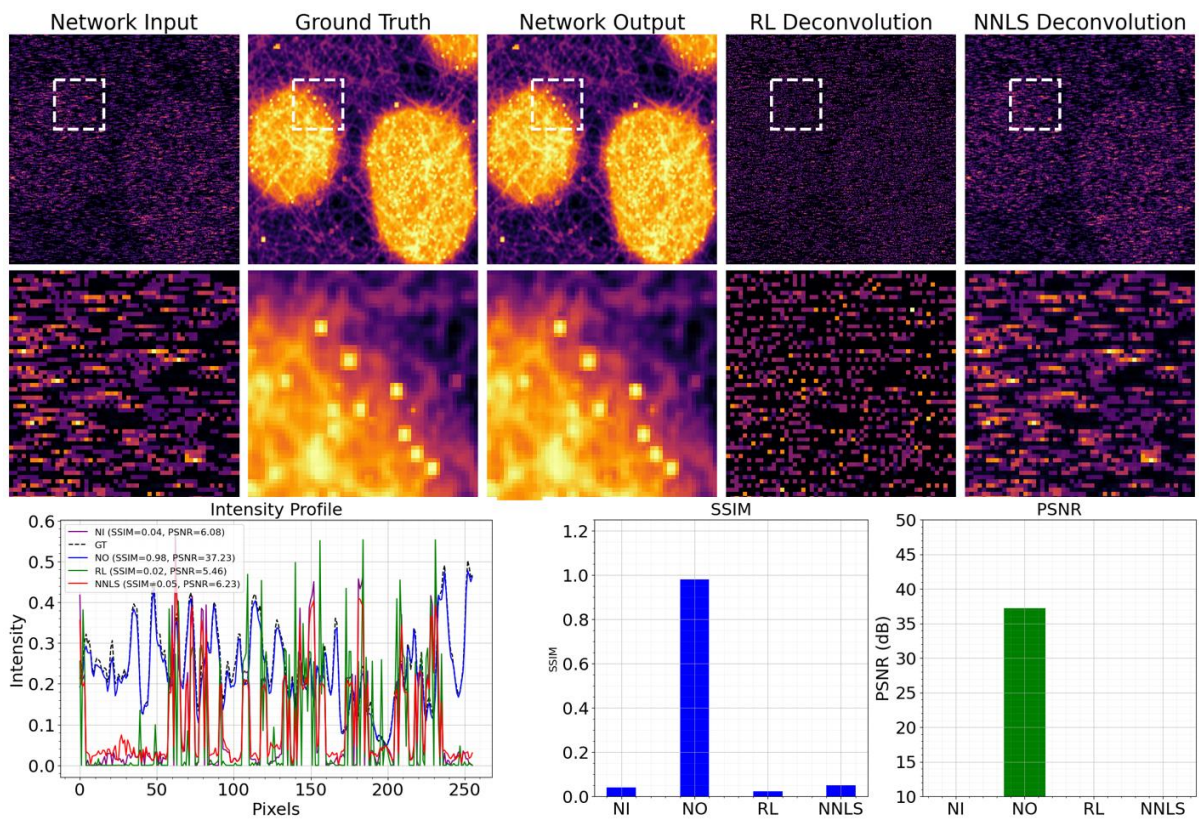

*Figure 5: Visualization of the recovery performance of the proposed Adaptive Physics Autoencoder for restoring degraded CLSM images of lipid droplets in a gel matrix.* The Network Input represents the noisy, degraded image, while the Ground Truth shows the high-resolution target. The Network Output demonstrates the reconstructed image compared to the results from the Richardson-Lucy (RL) and Non-Negative Least Squares (NNLS) deconvolution algorithms. Enlarged Regions of Interest (ROIs) highlight the superior restoration of droplet morphology and surrounding features by the proposed network. Quantitative evaluations using SSIM and PSNR further validate the performance improvement achieved by the proposed method.

The reconstructed confocal images of the network output, however, present a much finer structure than the deconvolution results. From the enlarged images of the white dotted line frame shown in the bottom row of Figure. 5, we can see that the network output image has reconstructed the circular droplet structures with a more obvious outline and the surrounding network structure closer to the real image.

Furthermore, we quantified the performance of images generated by different algorithms using SSIM and peak signal-to-noise ratio (PSNR) indexes. The quantitative results (bottom row) report that the SSIM of our algorithm is approximately 0.98, and the PSNR exceeds 36 dB. The experiment was repeated with 20 images, yielding similar results.

### 4.2 Performance on CLSM images of neuronal networks in cerebral organoids

The proposed Adaptive Physics Autoencoder was also applied to Confocal Laser Scanning Microscopy (CLSM) synthetic images of neuronal networks of cerebral organoids to assess its capability in reconstructing structurally complex and densely connected systems.

Figure 6 illustrates the comparative reconstruction performance of the network alongside traditional deconvolution methods. The network's output is directly inferred from degraded inputs, while the other approaches, Richardson-Lucy (RL) and non-negative least squares (NNLS) deconvolution algorithms, require preprocessing, including bilinear interpolation to match resolution requirements.

Unlike the RL and NNLS methods, which partially enhance the image resolution but struggle to reconstruct fine neuronal details, the network output excels in restoring the complexity of the neuronal architecture, including well-defined filaments and continuous network structures. Enlarged regions of interest (ROIs) in the bottom row further emphasize the network's ability to recover delicate connections and accurately represent the topology of neuronal systems.

Quantitative metrics support these visual observations, with the network achieving an SSIM of approximately 0.98 and a PSNR of 35.88 dB, significantly outperforming the deconvolution-based methods.

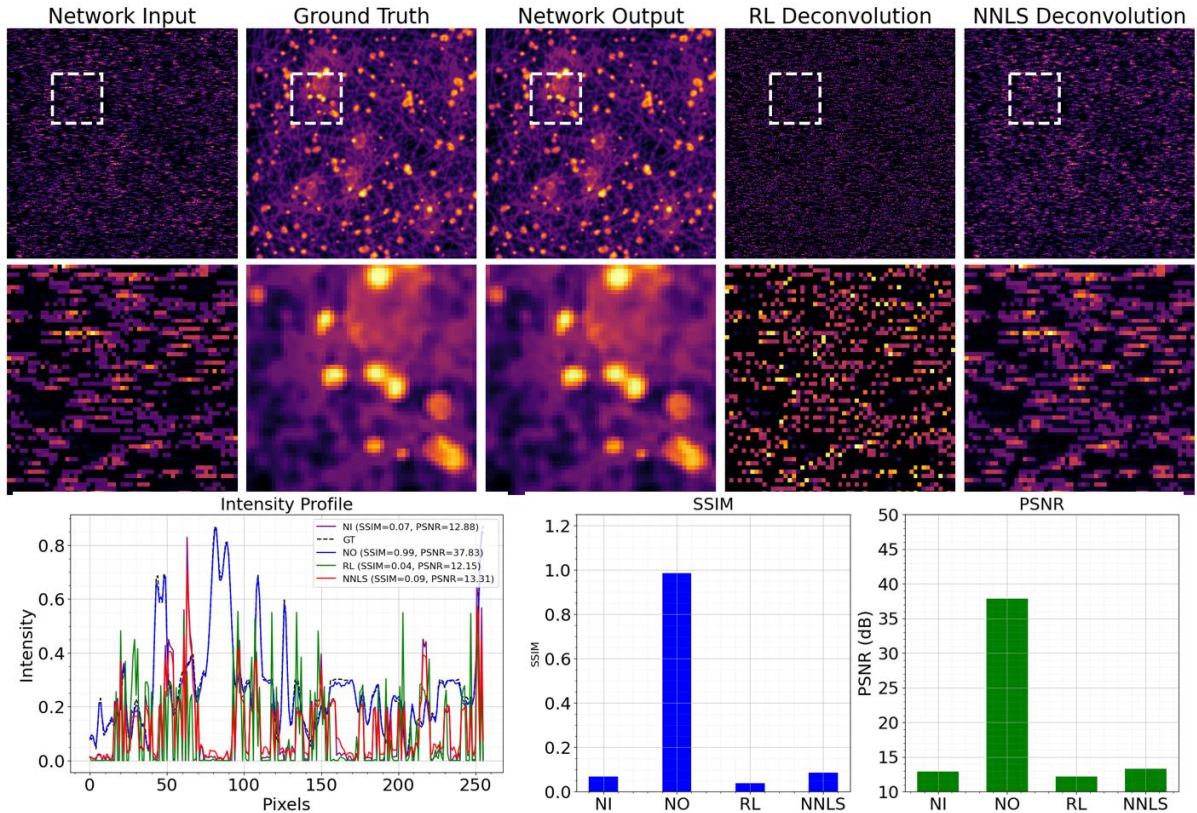

*Figure 6: **Confocal Laser Scanning Microscopy (CLSM) images of neuronal networks derived from cerebral organoids**, demonstrating the reconstruction performance of the proposed Adaptive Physics Autoencoder. The Network Input represents the degraded, noisy images, while the Network Output shows the restored images with improved structural fidelity. Comparisons with Richardson-Lucy (RL) and non-negative least squares (NNLS) deconvolution methods showcase the network's superior ability to recover fine neuronal details and maintain the continuity of the network. Enlarged Regions of Interest (ROIs) in 2$^{nd}$ row and quantitative metrics, including SSIM and PSNR, further showcase the improved performance of the proposed approach*

**4.3 Performance on CLSM images of sparse fibrillar structures**

The evaluation of the proposed Adaptive Physics Autoencoder was extended to synthetic CLSM images of sparse fibrillar structures. Figure 7 displays reconstruction performance of our model compared to traditional deconvolution methods. While RL and NNLS algorithms result in slight enhancements to the resolution of the input images, they struggle to accurately reconstruct the sparse fibrillar features. On the other hand, the network output demonstrates significant improvements, with well defined fibrillar structures and enhanced contrast closely resembling the Ground Truth.

Quantitative analysis also confirms these visual findings. The SSIM and PSNR values of the network output are markedly higher than those of RL and NNLS, with an SSIM of approximately 0.94 and a PSNR of 35.25 dB.

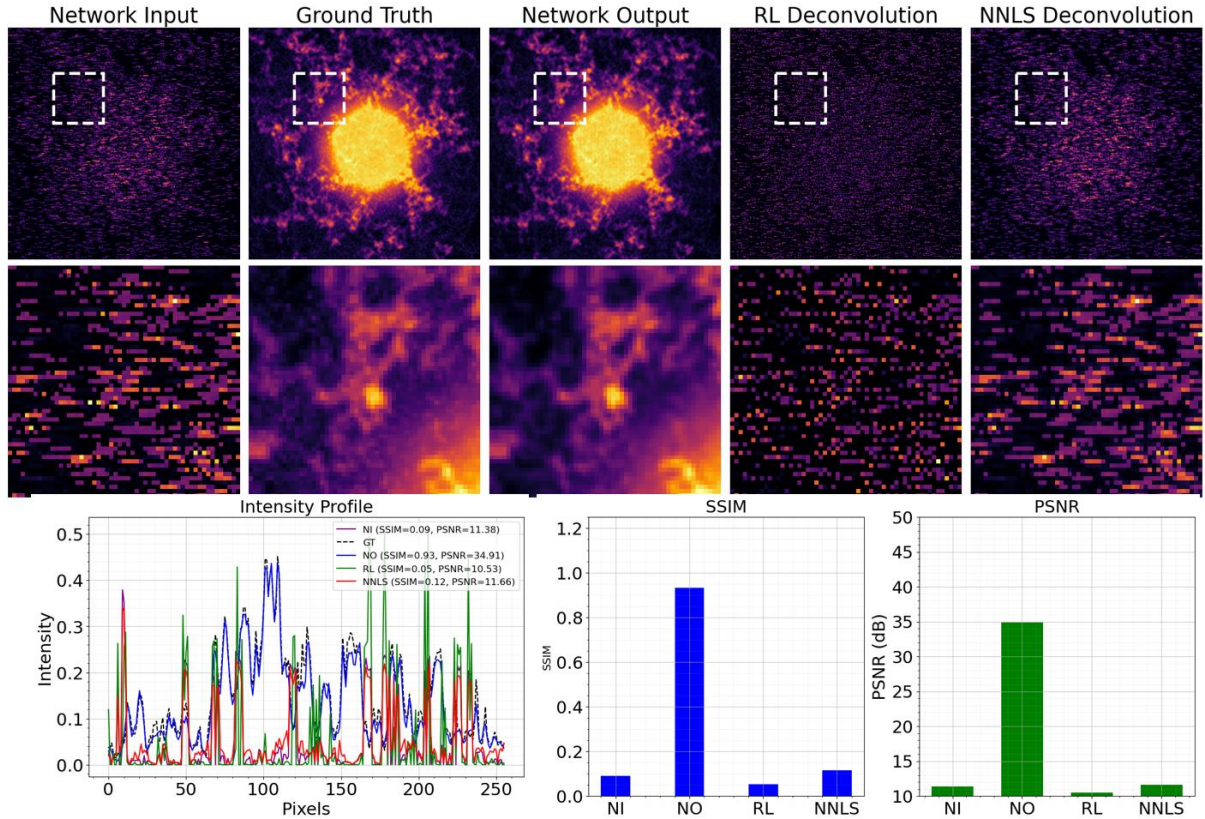

*Figure 7: Simulated CLSM images of sparse fibrillar structures. The Network Input represents the degraded images, while the Ground Truth shows the ideal high-resolution target. The Network Output highlights the network's ability to restore fine fibrillar details with high fidelity, compared to the results from Richardson-Lucy (RL) and non-negative least squares (NNLS) deconvolution methods. Enlarged Regions of Interest (ROIs) emphasize the superior restoration achieved by the network. Quantitative metrics, including SSIM and PSNR, demonstrate improved accuracy and structural preservation of the proposed approach*

### 4.4 Performance comparison with traditional reconstruction methods

The proposed Adaptive Physics Autoencoder is further evaluated by comparing its performance to additional image reconstruction methods, including Total Variation (TV) regularization, Wiener filtering, and Wavelet denoising, on simulated Confocal Laser Scanning Microscopy (CLSM) images of spherical structures embedded in a gel matrix, resembling microplastic particles. Figure 8 demonstrates the reconstruction performance of all methods using low-resolution, noisy inputs (Network Input) as the starting point, with the Ground Truth serving as the ideal reference.

While TV regularization, Wiener filtering, and Wavelet denoising improve the input image quality to some extent, they struggle to reconstruct fine details and introduce unwanted artifacts or excessive smoothing. On the other hand, the Network Output achieves better reconstruction of the spherical structures and preserves their boundaries and overall morphology closely matching the Ground Truth. Enlarged regions of interest (ROIs) show the superior performance of network in restoring high quality CLSM images.

The SSIM and PSNR values for the Network Output significantly outperform all other methods and achieved an SSIM of 0.88 and a PSNR of 32.82 dB. In contrast, TV regularization, Wiener filtering, and Wavelet denoising yield much lower SSIM and PSNR values, indicating their limited ability to handle the severe noise and undersampling present in the input images. The experiment confirms the network's robustness and superiority in restoring CLSM images across various scenarios.

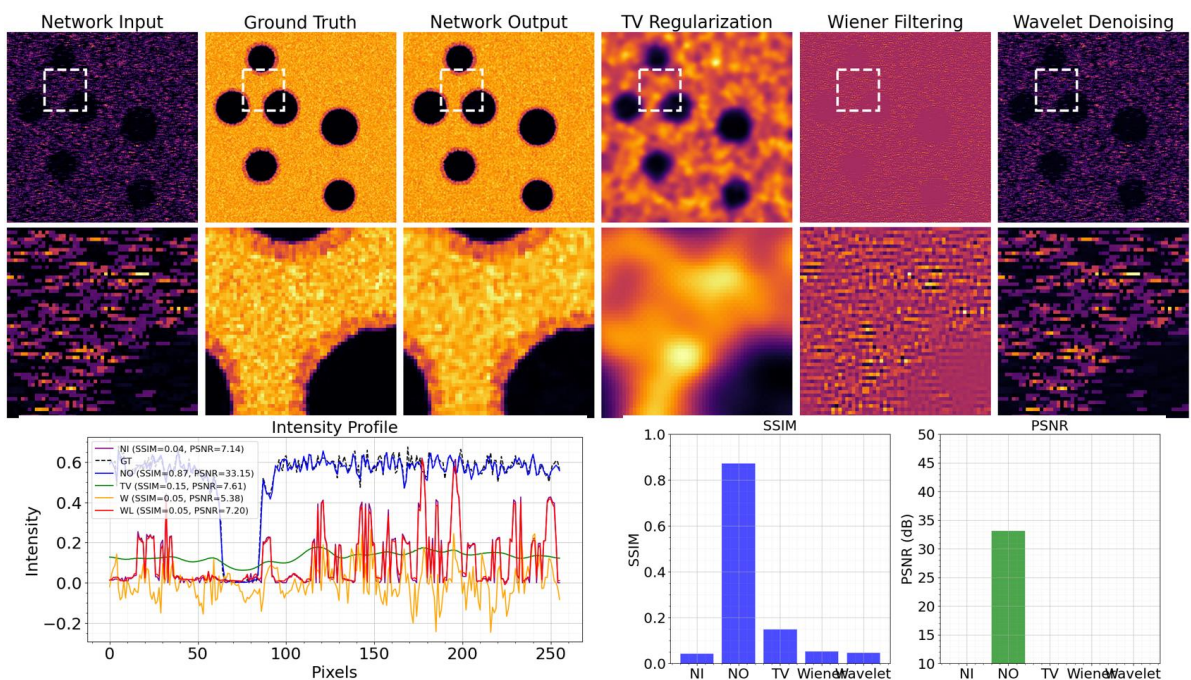

*Figure 8: Simulated Confocal Laser Scanning Microscopy (CLSM) images of spherical structures embedded in a gel matrix, resembling microplastic particles, comparing the reconstruction performance of the proposed Adaptive Physics Autoencoder with traditional methods. The Network Input represents the degraded, low-resolution image, while the Ground Truth provides the ideal reference. The Network Output demonstrates superior restoration quality, preserving structural details and boundaries, compared to Total Variation (TV) regularization, Wiener filtering, and Wavelet denoising. Enlarged Regions of Interest (ROIs) highlight the network's ability to achieve smooth, artifact-free reconstructions. Quantitative metrics, including SSIM and PSNR, validate the network's higher fidelity and noise suppression capability over the other methods.*

## 5. Conclusions

We proposed a deep learning-based method for high resolution image reconstruction for confocal microscopy. By simulating the real imaging process, our method reduces equipment costs, ensures consistency with real imaging conditions, and takes humans out of the loop, a step towards self-driving labs. The Adaptive Physics Autoencoder was evaluated extensively on various structures, including lipid droplets, cerebral neuronal networks, and fibrillar systems. Comparisons with widely used deconvolution algorithms and other reconstruction methods demonstrated its superior ability to recover fine structural details, validated both qualitatively and quantitatively through network output and ground truth comparison, SSIM, and PSNR metrics. In summary, Adaptive Physics Autoencoder represents a step forward in interpretable deep learning by integrating physics informed models with deep learning for confocal microscopy imaging, real time denoising of biological and medical imaging, and towards self-driving labs.

## 6. Future work

The current study focuses on imaging within the focal plane, future work will, however will be aimed at extending this method to volumetric imaging. and on real-time imaging systems. Other avenues include exploring lightweight network architectures and intensive physics constrained loss functions, enabling broader applications in High Resolution microscopy.

Furthermore, recent advancements, such as neural networks for aberration correction in optics systems and approaches combining convolutional neural networks with optical modeling show promise for more refinement of imaging models. Integrating this strategy into microscopy for real-time image denoising and enhancement can substantially enhance biological and medical imaging. The method could also be extended to areas such as histopathology, neuroanatomy, and other domains requiring precise imaging.

**Data availability statement**

The datasets in this paper are available upon reasonable request from the corresponding author. Specific details about the data sources and image preprocessing are described in the Methods section of this paper.

**GitHub codes availability statement**

The Python scripts developed and used in this study for performing image restoration are available on GitHub at [link]. It includes documentation and instructions for reproducing the results presented in this paper.